\documentclass[rnote]{aa} 
\usepackage{graphicx}
\usepackage{txfonts}
\usepackage{graphics}
\usepackage{epsf}

\def\ni{\noindent}

\begin{document}

   \title{Testing the nonlinearity of the $BVI_cJHK_s$ period-luminosity relations for the Large Magellanic Cloud Cepheids}

   \author{C. Ngeow
          \inst{1}
          \and
          S. M. Kanbur\inst{2}
	  \and
	  A. Nanthakumar\inst{3}
          }

   \offprints{C. Ngeow}

   \institute{Department of Astronomy, University of Illinois, Urbana, IL 61801, USA\\
     \email{cngeow@astro.uiuc.edu}
     \and
     Department of Physics, State University of New York at Oswego, Oswego, NY 13126, USA\\
     \email{kanbur@oswego.edu}
     \and
     Department of Mathematics, State University of New York at Oswego, Oswego, NY 13126, USA\\
     \email{nanthaku@oswego.edu}
   }
   
   \date{Received October 10, 2007; accepted October 24, 2007}

 
  \abstract
   {}
   {A number of recent works have suggested that the period-luminosity (PL) relation for the Large Magellanic Cloud (LMC) Cepheids exhibits a controversial nonlinear feature with a break period at 10 days. Therefore, the aim of this Research Note is to test the linearity/nonlinearity of the PL relations for the LMC Cepheids in $BVI_cJHK_s$ band, as well as in the Wesenheit functions.}
   {We show that simply comparing the long and short period slopes, together with their associated standard deviations, leads to a strictly larger error rate than applying rigorous statistical tests such as the $F$-test. We applied various statistical tests to the current published LMC Cepheid data. These statistical tests include the $F$-test, the testimator test, and the Schwarz information criterion (SIC) method.}
   {The results from these statistical tests strongly suggest that the LMC PL relation is nonlinear in $BVI_cJH$ band but linear in the $K_s$ band and in the Wesenheit functions. Using the properties of period-color relations at maximum light and multi-phase relations, we believe that the nonlinear PL relation is not caused by extinction errors. }
   {}


\titlerunning{Nonlinear PL Relations}
\authorrunning{Ngeow et al.}

   \maketitle

\section{Introduction}

Recently, Fouqu\'{e} et al. (\cite{fou07}) have derived the Galactic Cepheid period-luminosity (PL) relation with several different techniques, including parallax measurements (from {\it Hipparcos} and {\it HST}), variants of the Baade-Wesselink method, and distances inferred from open clusters. We point out that such an approach has been applied before in Ngeow \& Kanbur (\cite{nge04}) and Groenewegen et al. (\cite{gre04}). In addition, Fouqu\'{e} et al. (\cite{fou07}) also derive Large Magellanic Cloud (LMC) PL relations in the $BVR_cI_cJHK_s$ band, and refer to the work of Sandage et al. (\cite{san04}), which suggests a possible change of slope for the LMC PL relation at 10 days. In fact, there are several other papers on the topic of nonlinear\footnote{By nonlinearity we mean that the PL relation can be broken into two relations, with a break period adopted at 10 days.} LMC PL relations (see Kanbur \& Ngeow \cite{kan04,kan06}; Kanbur et al. \cite{kan07}; Ngeow et al. \cite{nge05}; Ngeow \& Kanbur \cite{nge06a,nge06b}; Koen et al. \cite{koe07}). 

These previous works concentrate on the $VI_c$ band (Kanbur \& Ngeow \cite{kan04,kan06}; Ngeow \& Kanbur \cite{nge06b}) or $V$ band only (Ngeow \& Kanbur \cite{nge06a}; Kanbur et al. \cite{kan07}), with data mostly from the OGLE (Optical Gravitational Lensing Survey, Udalski et al. \cite{uda99}) database. For the $JHK_s$ band PL relations, Ngeow et al. (\cite{nge05}) investigated possible nonlinearities using the 2MASS data from Nikolaev et al. (\cite{nik04}) that cross-correlated with the MACHO LMC Cepheids, and a random-phase correction to derive the mean magnitudes of these 2MASS data. 

Our motivation for this Research Note is to extend the previous work in $BVI_cJHK_s$ band, using the LMC Cepheid data from Fouqu\'{e} et al. (\cite{fou07}), with various rigorous statistical tests. The $JHK_s$ band data used in Fouqu\'{e} et al. (\cite{fou07}) are the 2MASS data matched to the OGLE Cepheids, and the mean magnitudes are derived using the method presented in Soszy\'{n}ski et al. (\cite{sos05}), which is different from the data used in Ngeow et al. (\cite{nge05}). It is important to test the nonlinearity results in $JHK_s$ band results with different Cepheid samples and different methods deriving the $JHK_s$ mean magnitudes. 

As emphasized in Ngeow \& Kanbur (\cite{nge06a}), statistical tests are needed to test and detect the existence of the nonlinear PL relation. We also point out that in searching for nonlinearity or a change of slope at 10 days, the method of comparing the short and long period slope with their associated standard deviations is more prone to error than applying a statistical test, such as the $F$-test as indicated by the following, purely analytical example. A statement such as the ``the slope is $x\pm \delta x$'' means that the probability that the slope is in the interval $(x-\delta x,x+\delta x)$ is $ 1 - \alpha$, where $\alpha$ is the desired significance level. Then if $A$ is the event that the calculated short period slope is wrong and $B$ is the event that the calculated long period slope is wrong, we have $P(A)=\alpha$ and $P(B) = \alpha$. Then in comparing the short and long period slopes using just their calculated standard deviations, the probability of at least one mistake is $P(A\cup B) = P(A) + P(B) - P(A\cap B) = 2{\alpha} - {\alpha}^2$. If $1 > {\alpha} > 0$, then $2{\alpha} - {\alpha}^2 > {\alpha}$. If the $F$-test or any other statistical test is carried out to the level of significance ${\alpha}$, then this states that the probability of making an error in just comparing long and short period slopes through their standard deviations is greater than the probability that the $F$-test makes a mistake. In essence the $F$-test compares both short and long period slopes (as well as the zero-points) of the nonlinear PL relations simultaneously.

\begin{table*}
\caption{$F$-test Results of the LMC PL Relations.}
\label{tab1}
\begin{center}
\begin{tabular}{lcccccccccc}
\hline\hline
Band & $a_S$ & $b_S$ & $\sigma_S$ & $N_S$ & $a_L$ & $b_L$ & $\sigma_L$ & $N_L$ & $F$ & $p(F)$ \\
\hline
$B$  & $-2.628\pm0.072$ & $17.493\pm0.046$ & 0.262 & 618 &  $-2.402\pm0.192$ & $17.419\pm0.238$ & 0.316 & 96 & 7.10 & 0.001 \\
$V$  & $-2.899\pm0.052$ & $17.148\pm0.033$ & 0.191 & 621 &  $-2.763\pm0.141$ & $17.127\pm0.176$ & 0.233 & 95 & 6.83 & 0.001 \\
$I_c$& $-3.073\pm0.035$ & $16.657\pm0.022$ & 0.126 & 604 &  $-2.951\pm0.104$ & $16.609\pm0.129$ & 0.162 & 88 & 7.15 & 0.001 \\
$J$  & $-3.237\pm0.040$ & $16.330\pm0.025$ & 0.126 & 481 &  $-3.035\pm0.151$ & $16.184\pm0.179$ & 0.134 & 48 & 5.00 & 0.007 \\
$H$  & $-3.347\pm0.036$ & $16.116\pm0.023$ & 0.114 & 481 &  $-3.099\pm0.137$ & $15.925\pm0.162$ & 0.122 & 48 & 7.69 & 0.001 \\
$K_s$& $-3.294\pm0.043$ & $16.027\pm0.028$ & 0.137 & 481 &  $-3.211\pm0.144$ & $15.992\pm0.171$ & 0.128 & 18 & 1.98 & 0.140 \\
\hline
$W_{bi}$  & $-3.463\pm0.021$ & $15.933\pm0.013$ & 0.074 & 598 &  $-3.507\pm0.055$ & $15.999\pm0.068$ & 0.086 & 88 & 0.886 & 0.413 \\
$W_{vi}$  & $-3.349\pm0.019$ & $15.897\pm0.012$ & 0.069 & 601 &  $-3.316\pm0.050$ & $15.883\pm0.062$ & 0.078 & 87 & 1.631 & 0.196 \\
\hline
\end{tabular} \\
\end{center}
The subscripts $_S$ and $_L$ are for the short ($\log P<1.0$) and long period Cepheids, respectively, while $a$, $b$ and $\sigma$ are the slope, zero-point and dispersion of the fitted PL relations.
\end{table*}

\section{Data \& Results from Statistical Tests}

The $BVI_cJHK_s$ band LMC Cepheid data were kindly provided by P. Fouqu\'{e}. It is {\it exactly} the same dataset used in Fouqu\'{e} et al. (\cite{fou07}). We do not include the $R_c$ band data because the number of Cepheids in $R_c$ band data is much smaller than in other band and some of them have a small number of data points per light curve. In addition to the $BVI_cJHK_s$ band, we also include the two Wesenheit functions considered in Fouqu\'{e} et al. (\cite{fou07}), namely $W_{bi}$ and $W_{vi}$. We fit a linear PL relation to these data and obtain identical PL relations as presented in table 8 of Fouqu\'{e} et al. (\cite{fou07}). The statistical tests we employed in this study include the $F$-test (Kanbur \& Ngeow \cite{kan04}; Ngeow et al. \cite{nge05}), the testimator test, and the Schwarz information criterion (SIC) method (Kanbur et al. \cite{kan07}). Details regarding the formalism and description for these statistical tests are given in the above references and will not be repeated here. 

\begin{figure}
\resizebox{\hsize}{!}{\includegraphics{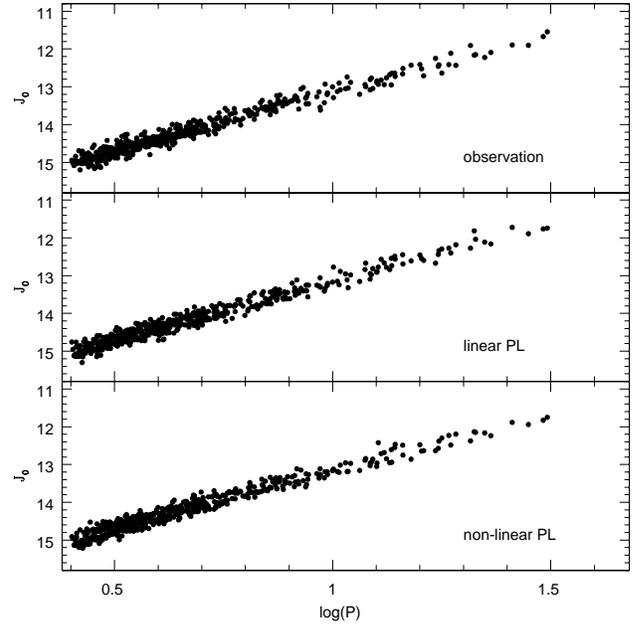}}
\caption{The top panel shows the $J$ band data from Fouqu\'{e} et al. (\cite{fou07}). Both of the middle and lower panels are from simulation. The simulation in the middle panel uses an intrinsic linear PL relation as input, while the PL relation in the bottom panel is simulated from an intrinsic nonlinear PL relation. The $F$-test as described in Section 2.1 returns $F=0.68$ and $F=7.62$ for the PL relations in the middle and bottom panels, respectively. The $F$-test correctly identified the PL relations that are intrinsically linear and nonlinear, respectively.}
\label{simu}
\end{figure}

To illustrate the difficulty of visualizing the nonlinear PL relation, if it truly exists, and the need for rigorous statistical tests to detect such nonlinearity, we simulate the $J$ band PL relation using the method detailed in Ngeow \& Kanbur (\cite{nge06a}). The input linear $J$ band PL relation is taken from  Fouqu\'{e} et al. (\cite{fou07}), and the input nonlinear PL relation in the simulation is adopted from Table \ref{tab1}. Figure \ref{simu} compares the $J$ band PL relation from the real data and the two simulations. The three PL relations in Figure \ref{simu} look similar and linear by eye. However the PL relation in the bottom panel is constructed from a nonlinear PL relation. As pointed out in Ngeow \& Kanbur (\cite{nge06a}), the difficulty of visualizing such a nonlinear PL relation is due to the existence of intrinsic dispersion of the PL relation caused by the finite width of the instability strip.

\subsection{The $F$-Test Results}

For the $F$-test, the null hypothesis ($H_0$) is a single regression line is sufficient, while the alternate hypothesis ($H_A$) is that two regression lines separated at 10 days are needed to fit the data. In Table \ref{tab1}, we present the results from the $F$-test, which include the fitted PL relations for the short ($\log P<1.0$) and long period Cepheids, the $F$ values and the probability, $p(F)$, under the null hypothesis. As in our previous work, the threshold for $p(F)$ was set to be $0.05$ (corresponds to 95\% confident level). For a large sample ($N>100$), $F\sim3$ at $p(F)=0.05$. Hence our $F$-test results indicate that the LMC PL relation is not linear in $BVI_cJH$ band but linear in $K_sW_{bi}W_{vi}$ band. Note that some of the slopes and/or zero-points in Table \ref{tab1} appear to be consistent between the short and long period Cepheids, but this does not negate the nonlinearity of the PL relation as pointed out in the Introduction.

\begin{figure*}
  \vspace{0cm}
  \centering
    \epsfxsize=8.0cm \epsfbox{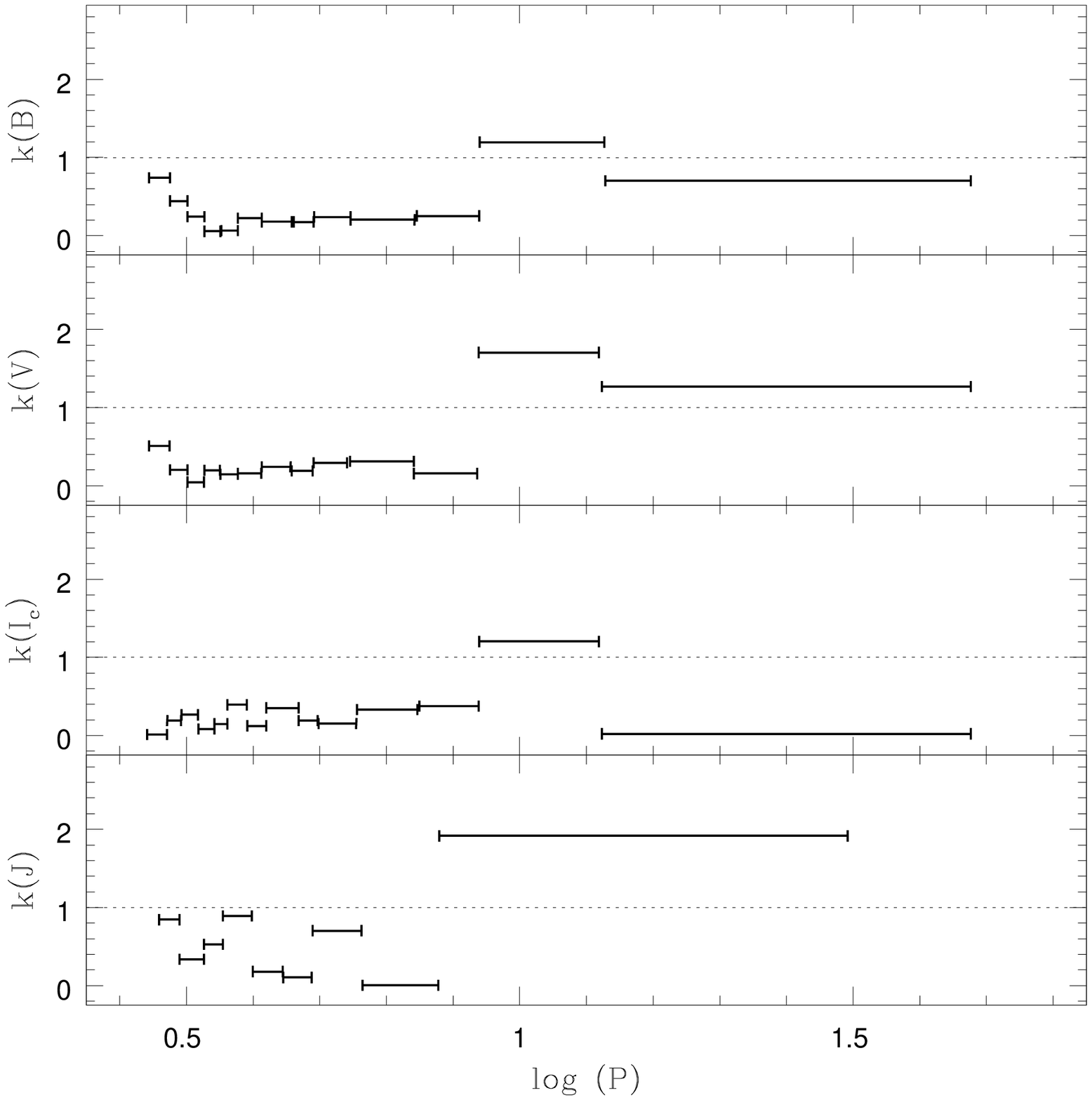}
    \epsfxsize=8.0cm \epsfbox{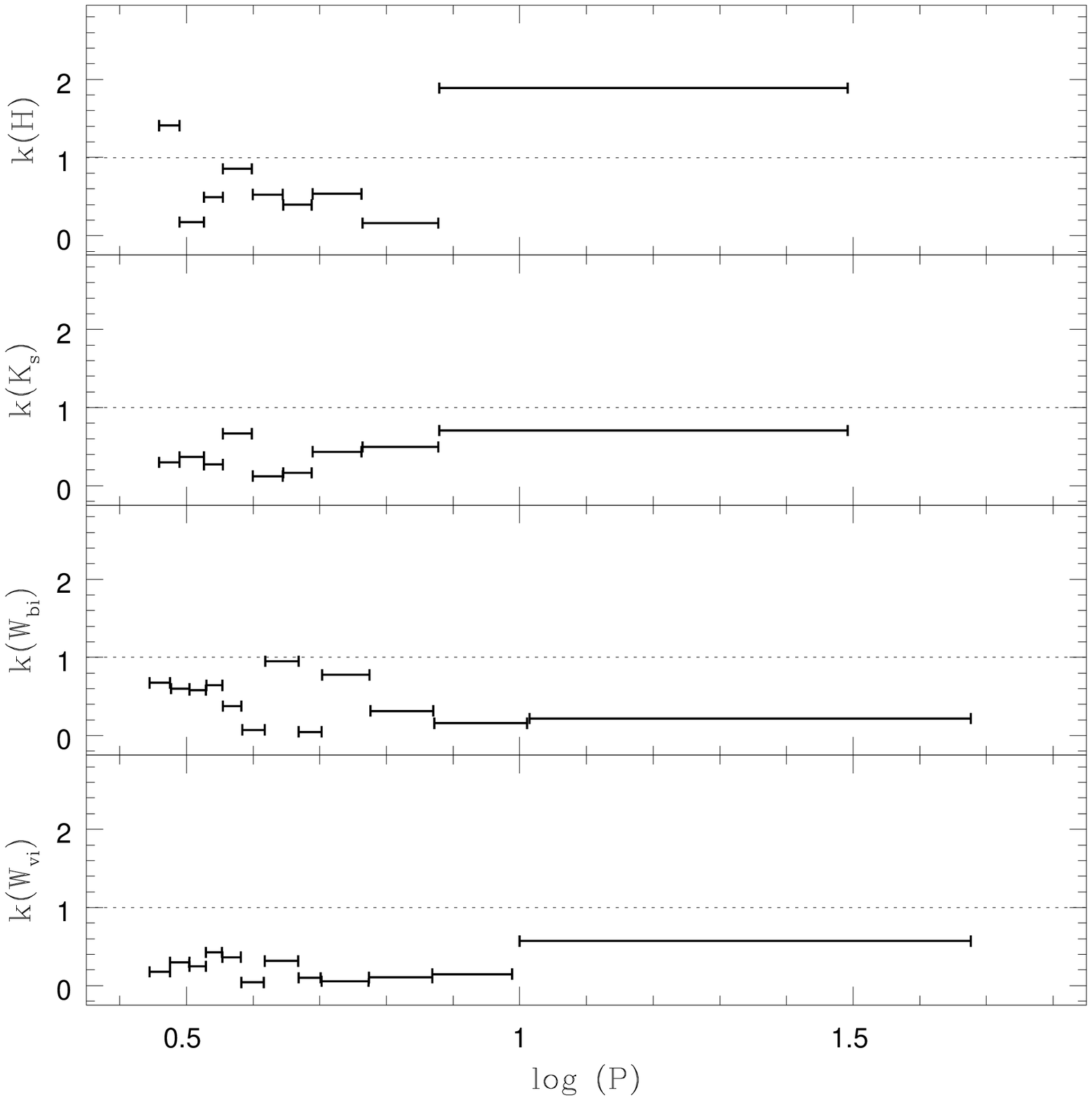}
  \vspace{0cm}
  \caption{The results from the testimator test. The horizontal bars are the $k$ values for each sub-samples. The size of the bars indicates the period range covered in each sub-samples. The number of data points in each sub-samples ranges from 45 to 88. The dashed lines are for the case $k=1$. \label{fig2}}
  \end{figure*}

\begin{figure*}
  \vspace{0cm}
  \centering
    \epsfxsize=8.0cm \epsfbox{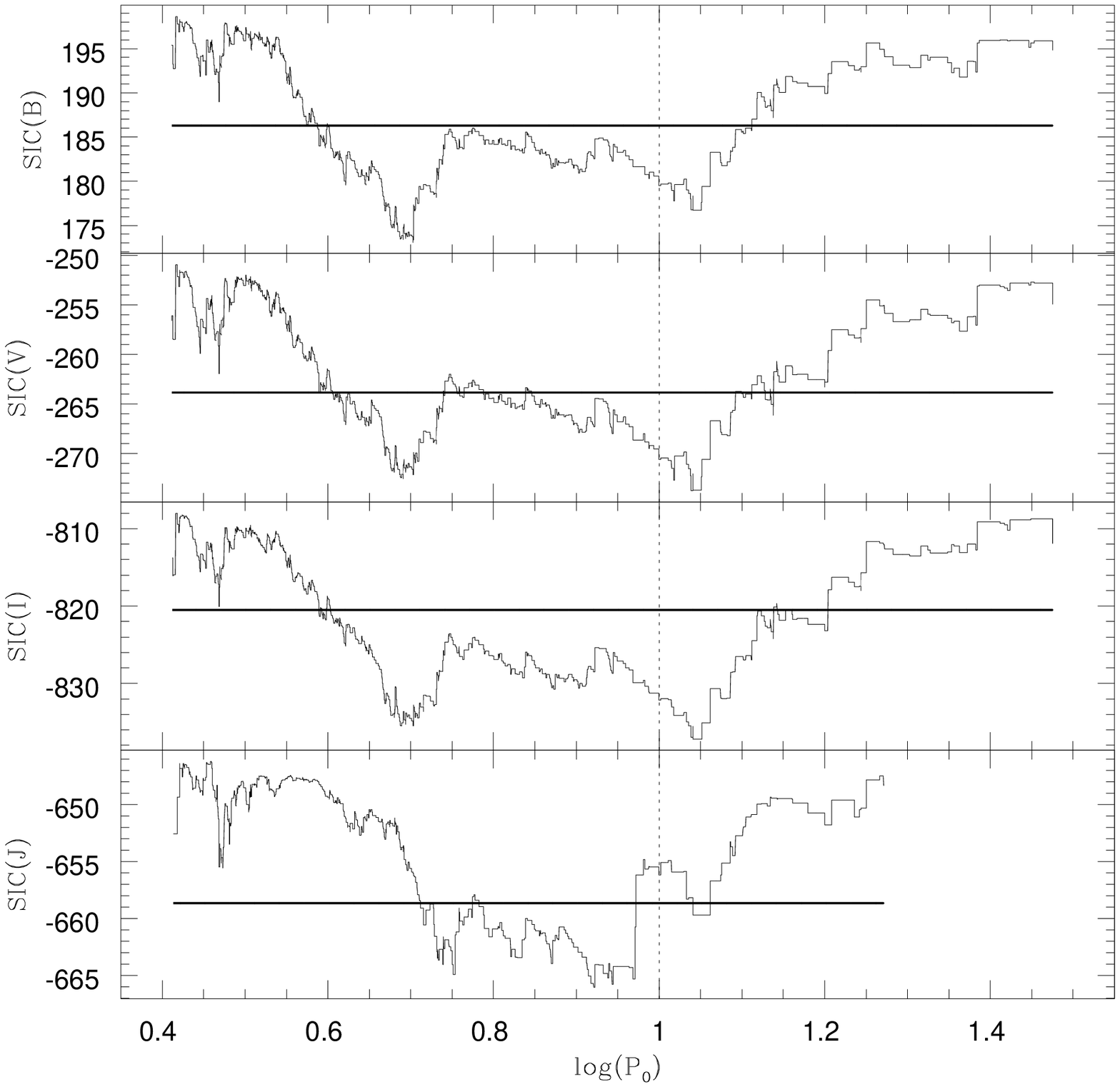}
    \epsfxsize=8.0cm \epsfbox{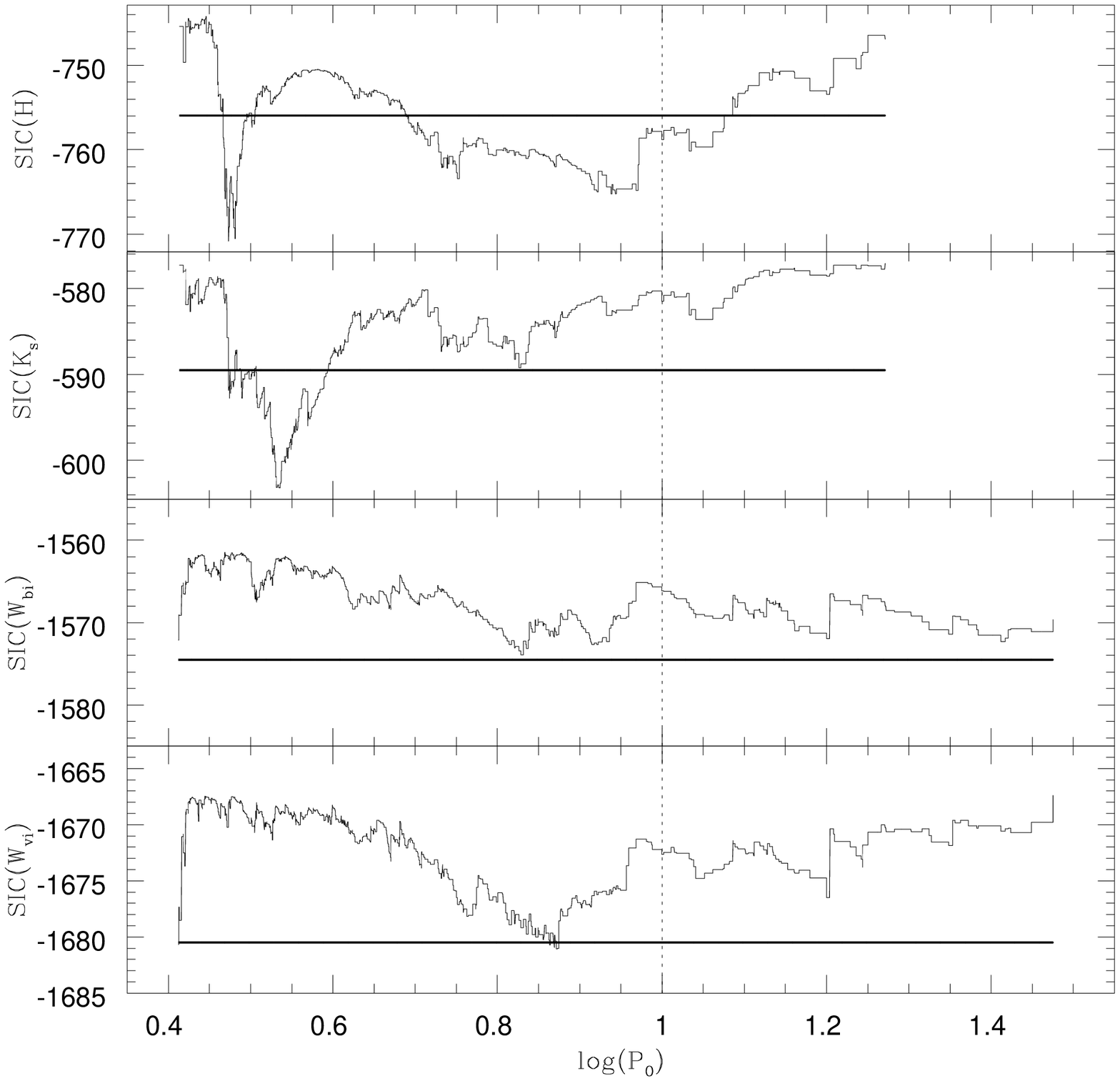}
  \vspace{0cm}
  \caption{The results from the SIC test. The thick horizontal lines (since it is independent of $\log P_0$) and the ``curves'' are for the null and alternate hypothesis, respectively. The vertical dashed lines indicate the adopted break period at 10 days. \label{fig3}}
  \end{figure*}

\subsection{The Testimator-Test Results}

Briefly, the testimator test requires the Cepheid sample to be divided into a number of sub-samples, after the sample has been sorted according to the periods. The slope of each sub-sample is compared to the slope from the previous sub-sample (except the first sub-sample) using the $t$ statistical test. The null hypothesis for the $t$-test is that the two slopes are statistically equal, and the alternate hypothesis is that they are not. If the null hypothesis is rejected, a new slope, called the testimator, is calculated. The ratio, $k$, of the observed $t$ values and the critical $t$ value is calculated for each sub-sample (see Kanbur et al. \cite{kan07} for more details). The case of $k>1$ indicates that the null hypothesis can be rejected for a given sub-sample, and the slope for this sub-sample is statistically different to the slope in the previous sub-sample. This ensures that the testimator tests only the slope and not differences in zero point. In Figure \ref{fig2}, we present the $k$ values of each of the sub-samples under the testimator test. From the figure, it can be seem that in the $BVI_cJH$ band, the slopes change (with $k>1$) for sub-samples that bracket the assumed break period at 10 days and/or the sub-samples with longer period Cepheids. In contrast, the slopes do not change statistically in the sub-samples for $K_sW_{bi}W_{vi}$ band.

\subsection{The SIC-Test Results}

For the SIC test, the null hypothesis is taken to be a linear regression model, while the alternate hypothesis is a nonlinear regression model with a break period at $P_0$. This break period is varied over the entire period range and likelihoods under the null and alternate hypotheses are calculated. We look for values of $P_0$ for which the likelihood under the alternative hypothesis is greater than that under the null hypothesis. The model with the lowest $SIC$ value is the preferred model. Figure \ref{fig3} summarizes the results from this SIC test. In $BVI_cJH$ band, the SIC test finds evidence that there is a range of $P_O$ where the alternate hypothesis is a preferred model. This range of $P_O$ includes the adopted break period at 10 days (except in the $J$ band, however the range of the break period is still close to 10 days). The existence of a range of $P_0$ and the difficulty of pin-pointing the break period in the SIC test is mainly due to the finite width of the instability strip (see more detailed discussion in Kanbur et al. \cite{kan07}). The $K_s$ band results do not show any preferred alternate models around 10 days. The SIC results for the two Wesenheit functions also do not prefer the alternate hypotheses across all $P_0$. Interestingly, the $HK_s$ band imply that at $\log P_0\sim0.5$ the alternate hypothesis show a smaller value of SIC than the null hypothesis. Currently there is no explanation to account for this.

\subsection{Tests for the $JHK_s$ Band PL Relations with Additional Data}

\begin{table*}
\caption{$F$-test Results of the $JHK_s$ PL Relations with additional data from Persson et al. (\cite{per04}).}
\label{tab2}
\begin{center}
\begin{tabular}{lcccccccccc}
\hline\hline
Band & $a_S$ & $b_S$ & $\sigma_S$ & $N_S$ & $a_L$ & $b_L$ & $\sigma_L$ & $N_L$ & $F$ & $p(F)$ \\
\hline
$J$  & $-3.234\pm0.038$ & $16.328\pm0.024$ & 0.125 & 499 & $-3.255\pm0.077$ & $16.416\pm0.098$ & 0.144 & 116 & 7.29 & 0.001 \\
$H$  & $-3.343\pm0.034$ & $16.114\pm0.022$ & 0.113 & 499 & $-3.300\pm0.066$ & $16.158\pm0.084$ & 0.123 & 116 & 8.44 & 0.000 \\
$K_s$& $-3.300\pm0.041$ & $16.030\pm0.026$ & 0.135 & 499 & $-3.371\pm0.065$ & $16.169\pm0.083$ & 0.121 & 116 & 3.03 & 0.049 \\
\hline
\end{tabular}
\end{center}
The symbols are same as in Table \ref{tab1}.
\end{table*}

\begin{figure*}
  \vspace{0cm}
  \centering
    \epsfxsize=8.0cm \epsfbox{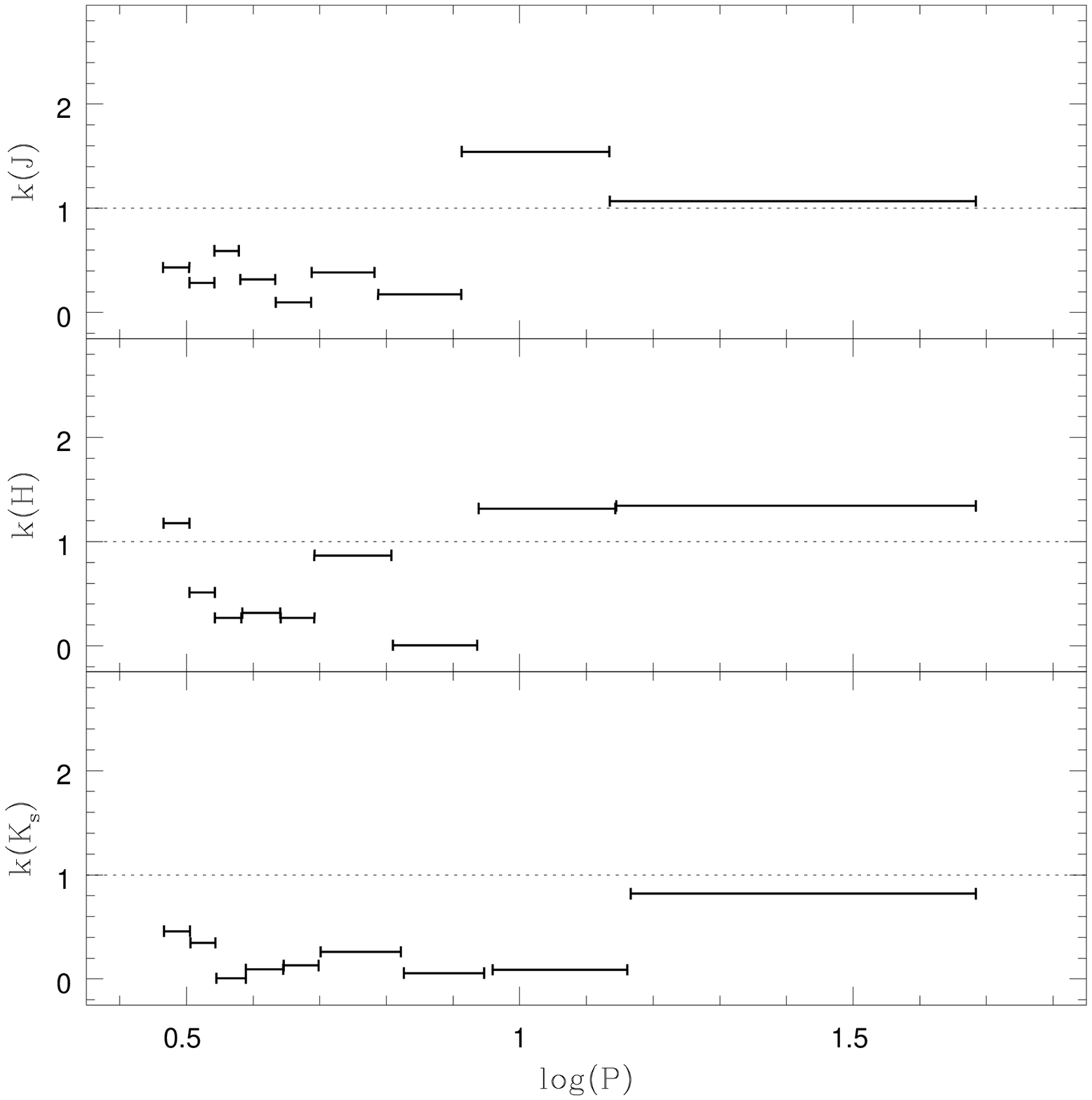}
    \epsfxsize=8.0cm \epsfbox{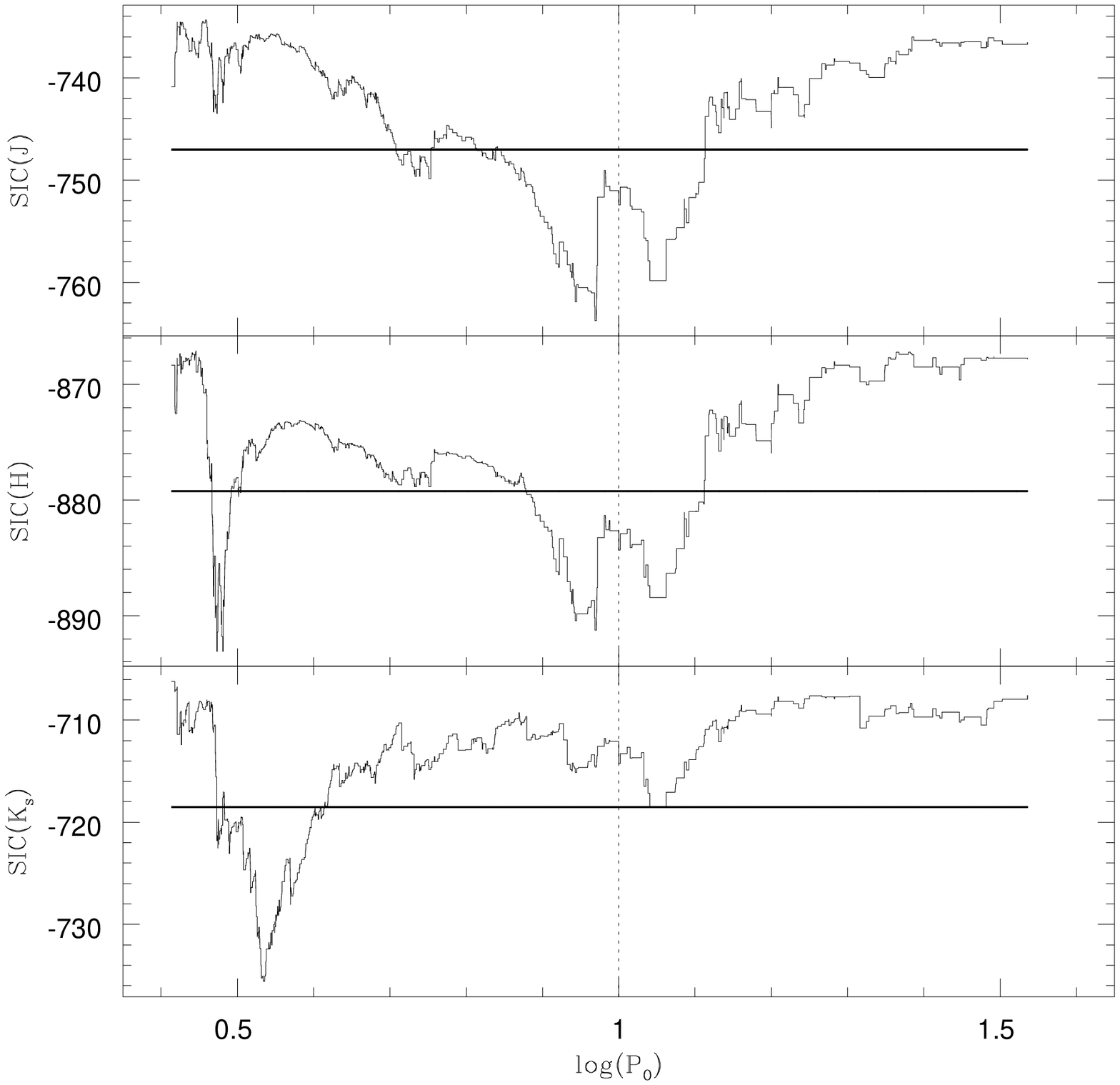}
  \vspace{0cm}
  \caption{Test results for the testimator test (left panel) and the SIC test (right panel) for the $JHK_s$ band PL relations with additional data from Persson et al. (\cite{per04}). See Figure \ref{fig2} and \ref{fig3} for the meaning of the lines and curves.  \label{fig4}}
  \end{figure*}

In contrast to other band, the $JHK_s$ data from Fouqu\'{e} et al. (\cite{fou07}) only consist of the OGLE Cepheids. Therefore, we include additional $JHK_s$ band Cepheid data from Persson et al. (\cite{per04}) to the sample, as suggested by the referee. As in Kanbur \& Ngeow (\cite{kan06}) and Fouqu\'{e} et al. (\cite{fou07}), we only include Cepheids with $\log (P)<1.8$ in the sample. Further we remove Cepheid HV 12765 from the sample as suggested in Persson et al. (\cite{per04}). Since the $JHK_s$ data from Fouqu\'{e} et al. (\cite{fou07}) is in 2MASS system and the Persson et al. (\cite{per04}) $JHK_s$ data is in LCO system, we convert the Persson et al. (\cite{per04}) $JHK_s$ data to the 2MASS system using a shift of $-0.02$mag. as stated in Fouqu\'{e} et al. (\cite{fou07}). After applying the extinction correction, the linear $JHK_s$ PL relations with the combined $615$ Cepheids are:

\begin{eqnarray}
m(J)   & = & -3.121(\pm0.018)\log (P) + 16.263(\pm0.014), \nonumber \\
m(H)   & = & -3.228(\pm0.016)\log (P) + 16.047(\pm0.013), \nonumber \\
m(K_s) & = & -3.249(\pm0.018)\log (P) + 16.001(\pm0.015),\nonumber 
\end{eqnarray}

\ni with dispersion of $0.130$, $0.117$ and $0.133$ respectively.

The $F$-test results for the combined $JHK_s$ data is presented in Table \ref{tab2}, while the results from the testimator test and the SIC test are collectively summarized in Figure \ref{fig4}. All three statistical tests again found strong evidence of nonlinear $JH$ band PL relations, at the assumed break period of 10 days, for the combined Cepheid data. For the $K_s$ band, the $F$-test shows that the PL relation is marginally linear, with the support from the testimator and SIC tests that the nonlinearity is not detected at the break period of 10 days.

\section{Conclusion \& Discussion}

Combining the results from the three statistical tests presented in the previous sections, we find that there is strong statistical evidence to suggest the LMC PL relation is nonlinear in the $BVI_cJH$ band but linear in the $K_sW_{bi}W_{vi}$ band. Including additional data from Persson et al. (\cite{per04}) for the $JHK_s$ band does not alter the results as well. We have to emphasize that both of the testimator and SIC methods are applied to the $BI_cJHK_s$ band and the Wesenheit functions for the first time, in contrast to the $V$ band data that has been studied in Kanbur et al. (\cite{kan07}).

The nonlinear LMC PL relation has been found from a Cepheid sample that consists of OGLE Cepheids only (Kanbur \& Ngeow \cite{kan04}). To extend the OGLE sample, mostly at the long period end, Ngeow \& Kanbur (\cite{nge06a}) included various additional data from literature (see table 1 of Ngeow \& Kanbur \cite{nge06a}), and again found strong evidence of nonlinearity of the LMC PL relation. In this Research Note, results using the Fouqu\'{e} et al. (\cite{fou07}) data alone, and with additional data from Persson et al. (\cite{per04}), further supports the conclusion given in Ngeow \& Kanbur (\cite{nge06a}): that the sample selection does not play an important role in detecting the nonlinear LMC PL relation. However, Fouqu\'{e} et al. (\cite{fou07}) have suggested that the mixture of data used in previous work may lead to the nonlinearity seen in the statistical tests. This may certainly be the case and the analysis of a homogeneous sample, such as that provided by the ``LMC shallow survey'' (Fouqu\'{e} 2007; Gieren 2007 -- private communication) is desirable.

The nonlinearity of the PL relation that is seen in the optical and $JH$ band but not in the reddening insensitive $K_s$ band and the Wesenheit function may suggest that extinction is the cause of the nonlinearity. However, extinction is not the only explanation and there is some evidence against the hypothesis of extinction errors as a cause for the apparent nonlinearity. The linearity of the $K_s$ band PL relation, as compared to other shorter wavelength PL relations, is expected from black-body arguments (Ngeow \& Kanbur \cite{nge06a}). Simply speaking, the temperature variation dominates the luminosity variation in the optical, and extends to $JH$ band for Cepheid-like temperatures. But in the $K_s$ band the luminosity variation is dominated by the radius variation of Cepheid variables. The proposed mechanism that may cause the nonlinear PL relation, the interaction between the hydrogen ionization front and the stellar photosphere (Kanbur \& Ngeow \cite{kan06}), will only affect the temperature variation and not the radius variation. The linearity of the Wesenheit functions is also not a surprise, and has been studied and discussed in Ngeow \& Kanbur (\cite{nge05w}) and in Koen et al. (\cite{koe07}), and will not be repeated here. 

\begin{figure}
\resizebox{\hsize}{!}{\includegraphics{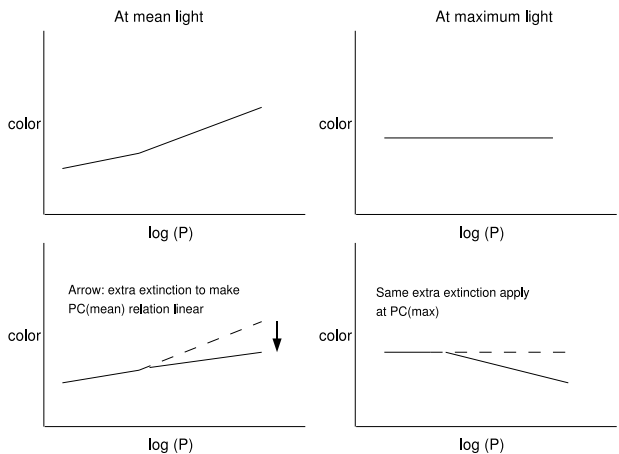}}
\caption{Schematic illustration for the argument with PC(max) relation. Top panels show the observed PC relations (after corrected for extinction) at mean (top-left panel) and maximum (top-right panel) light. Bottom panels show that if additional extinction as function of period to make the mean light PC relation linear, then the same extinction will cause the colors at maximum light get bluer as period increases, which are against observation and theoretical expectation.}
\label{pcmax}
\end{figure}

Since the additional data used is mainly at the long period end, the possibility remains of systematic errors in reddening as a function of period. However we note that a reddening error as a function of long period LMC Cepheids would also change the observed properties of LMC Cepheids at other phases. A reddening error as a function of period such that LMC Cepheids obey a linear PL relation at mean light would force the LMC Cepheids to have a period-color relation such that they get bluer or hotter at maximum light as the period increases (see Figure \ref{pcmax} for a schematic illustration). This is in stark contrast to the behavior of Galactic Cepheids and long period LMC Cepheids, which are known to have a flat period-color relation at maximum light (Code \cite{cod47}; Simon et al \cite{sim93}; Kanbur \& Ngeow \cite{kan04,kan06}; Kanbur et al \cite{kan04a}). Moreover, it is difficult to explain, theoretically, how a Cepheid could get hotter at maximum light as the period increases. 

The PL relation at phase $\sim0.8$ described in Ngeow \& Kanbur (\cite{nge06b}) presents clearly the dramatic nature of the nonlinearity at 10 days and the dynamic nature of the PL relation as a function of phase. It is difficult to reconcile this behavior as being due to sampling errors and/or reddening errors. It is worth to point out that the mean light PL relation used in the literature is an average of the PL relations in all phases (Kanbur \& Ngeow \cite{kan04}; Kanbur et al. \cite{kan04a}; Ngeow \& Kanbur \cite{nge06b}). nonlinearity of the PL relations at certain phases will certainly affect the linearity/nonlinearity of the mean light PL relation.

We have to remind that the data used in this study (and in most of our previous work) were published data that have been corrected for extinction using the ``state-of-the-art'' and well-developed methodology. If extinction error is believed to be the cause of nonlinear PL relations, it would imply that the extinction correction done previously in the literature is incorrect and/or incomplete. This would affect the previous work that using these extinction corrections, and those results need to be revised in future work.

\begin{acknowledgements}
We would like to thank the referee, P. Fouqu\'{e}, for useful discussion. We also thank P. Fouqu\'{e} and I. Soszy\'{n}ski for sharing their data. CN acknowledges support from NSF award OPP-0130612 and a University of Illinois seed funding award to the Dark Energy Survey. SMK acknowledges support from the Chretien International Research Award from the American Astronomical Society. 
 
\end{acknowledgements}


\begin{thebibliography}{}
\bibitem[1947]{cod47} Code A. D., 1947, ApJ, 106, 309

\bibitem[2007]{fou07} Fouqu\'{e}, P., Arriagada, P., Storm, J., et al., 2007, A\&A in press (arXiv:0709.3255)

\bibitem[2004]{gre04} Groenewegen, M. A. T., Romaniello, M., Primas, F. \& Mottini, M., 2004, A\&A, 420, 655

\bibitem[2004]{kan04} Kanbur, S. \& Ngeow, C., 2004, MNRAS, 350, 962 

\bibitem[2004]{kan04a} Kanbur, S., Ngeow, C. \& Buchler, R., 2004, MNRAS, 354, 212 

\bibitem[2006]{kan06} Kanbur, S. \& Ngeow, C., 2006, MNRAS, 369, 705 

\bibitem[2007]{kan07} Kanbur, S., Ngeow, C., Nanthakumar, A. \& Stevens, R., 2007, PASP, 119, 512

\bibitem[2007]{koe07} Koen, C., Kanbur, S. \& Ngeow, C., 2007, MNRAS, 380, 1440

\bibitem[2004]{nge04} Ngeow, C. \& Kanbur, S., 2004, MNRAS, 349, 1130

\bibitem[2005]{nge05w} Ngeow, C. \& Kanbur, S., 2005, MNRAS, 360, 1033

\bibitem[2005]{nge05} Ngeow, C., Kanbur, S., Nikolaev, S., et al., 2005, MNRAS, 363, 831

\bibitem[2006a]{nge06a} Ngeow, C. \& Kanbur, S., 2006a, ApJ, 650, 180

\bibitem[2006b]{nge06b} Ngeow, C. \& Kanbur, S., 2006b, MNRAS, 369, 723

\bibitem[2004]{nik04} Nikolaev, S., Drake, A. J., Keller, S. C., et al., 2004, ApJ, 601, 260

\bibitem[2004]{per04} Persson, S., Madore, B., Krzemi\'{n}ski, W., Freedman., W., Roth, M. \& Murphy, D., 2004, AJ, 128, 2239

\bibitem[2004]{san04} Sandage, A., Tammann, G. A. \& Reindl, B., 2004, A\&A, 424, 43

\bibitem[1993]{sim93} Simon, N., Kanbur, S. \& Mihalas, D., 1993, ApJ, 414, 310 

\bibitem[2005]{sos05} Soszy\'{n}ski, I., Gieren, W. \& Pietrzy\'{n}ski, G., 2005, PASP, 117, 823

\bibitem[1999]{uda99} Udalski, A., Soszy\'{n}ski, I., Szymanski, M., et al., 1999, Acta Astron., 49, 223 
\end{thebibliography}
\end{document}